\documentclass[10pt, conference, letterpaper]{IEEEtran}
\renewcommand{\baselinestretch}{1}
\usepackage{subfigure}
\usepackage[font=small,labelfont=bf]{caption}

\usepackage{color}
\usepackage{graphicx}
\usepackage{booktabs} 
\usepackage{listings}
\usepackage{url}
\usepackage{threeparttable}
\newcommand{\tabincell}[2]{\begin{tabular}{@{}#1@{}}#2\end{tabular}}

\usepackage[ruled]{algorithm2e} 

\usepackage[belowskip=-5.5pt,aboveskip=-0.5pt]{caption} 
\renewcommand{\baselinestretch}{.983}
\newcommand{\ignore}[1]{}

\definecolor{dkgreen}{rgb}{0,0.6,0}
\definecolor{gray}{rgb}{0.5,0.5,0.5}
\definecolor{mauve}{rgb}{0.58,0,0.82}

\begin{document}

\title{Keeping Context In Mind: Automating Mobile App Access Control with User Interface Inspection}

\author{
    \IEEEauthorblockN{
    Hao Fu\IEEEauthorrefmark{1},
    Zizhan Zheng\IEEEauthorrefmark{2},
    Sencun Zhu\IEEEauthorrefmark{3},
    Prasant Mohapatra\IEEEauthorrefmark{1}}
    \IEEEauthorblockA{\IEEEauthorrefmark{1}Department of Computer Science, University of California, Davis, USA.}\vspace{-3.5mm} \\
    \IEEEauthorblockA{\IEEEauthorrefmark{2}Department of Computer Science, Tulane University, New Orleans, USA.}\vspace{-3.5mm} \\
    \IEEEauthorblockA{\IEEEauthorrefmark{2}Department of Computer Science, Pennsylvania State University, University Park, USA.}\vspace{-3.5mm} \\
    \texttt{\{haofu, pmohapatra\}@ucdavis.edu, zzheng3@tulane.edu, szhu@cse.psu.edu}
}

\maketitle
\begin{abstract}
Recent studies observe that app foreground is the most striking component that influences the access control decisions in mobile platform, as users tend to deny permission requests lacking visible evidence.
However, none of the existing permission models provides a systematic approach that can automatically answer the question: \emph{Is the resource access indicated by app foreground?}

In this work, we present the design, implementation, and evaluation of {\tt COSMOS}, a context-aware mediation system that bridges the semantic gap between foreground interaction and background access, in order to protect system integrity and user privacy. 
Specifically, {\tt COSMOS} learns from a large set of apps with similar functionalities and user interfaces to construct generic models that detect the outliers at runtime.
It can be further customized to satisfy specific user privacy preference by continuously evolving with user decisions.
Experiments show that {\tt COSMOS} achieves both high precision and high recall in detecting malicious requests. 
We also demonstrate the effectiveness of {\tt COSMOS} in capturing specific user preferences using the decisions collected from 24 users and illustrate that {\tt COSMOS} can be easily deployed on smartphones as a real-time guard with a very low performance overhead.

\ignore{
When a security or privacy sensitive behavior is triggered, 
{\tt INSPIRED} tries to answer the following three questions: \emph{who} initiated the behavior, \emph{when} was the sensitive action triggered and under \emph{what} kind of environment was it triggered? Based on the answers, it then determines whether to grant the relevant permission or not. 
  Specifically, 
    we propose a machine learning based permission model using foreground information obtained from multiple sources.
    To precisely capture user intention, our permission model evolves over time and it can be user-customized by continuously learning from user decisions.
}

\end{abstract}

\section{Introduction}\label{sec:intro}
Mobile operating systems such as Android and iOS adopt permission systems that 
allow users to grant or deny a permission request when it is needed by an app for the first time.
But this approach does not provide sufficient protection as an adversary can easily induce users to grant the permission first, and then exploit the same resource for malicious purposes.
A recent user study~\cite{wagner2015} showed that at least 80\% users would have preferred to preventing at least one permission request involved in the study and suggested the necessity of more fine-grained control of permissions.
Ideally, a permission system should be able to identify suspicious permission requests {\it on the fly} and {\it automatically} by taking user preferences into account and notify users only when necessary.
As shown in several user studies~\cite{wagner2015,wagner2017,smarper}, it is crucial to consider the {\it context} pertinent to sensitive permission requests. Moreover, a user's preference is strongly correlated with the foreground app and the visibility of the permission requesting app (i.e., whether the app is currently visible to the user).
The intuition is that users often rely on displayed information to infer the purpose of a permission request and they tend to block requests that are considered to be irrelevant to app's functionalities~\cite{wagner2015}.
Thus, a permission system that can properly identify and utilize foreground data may significantly improve decision accuracy and reduce user involvement.
We posit that to fully achieve {\it contextual integrity}~\cite{contextualintegrity},
it is crucial to capture detailed foreground information 
by inspecting \textbf{\emph{who}} is requesting the permission, \textbf{\emph{when}} the request is initiated, and under \textbf{\emph{what}} circumstances it is initiated, in order to model the precise context surrounding a request.

In this paper,
we present the design and implementation of a lightweight run-time permission control system named {\tt COSMOS} (COntext-Sensitive perMissiOn System).
{\tt COSMOS} detects unexpected permission requests through examination of contextual foreground data.
For instance, a user interacting with an SMS composing page would expect the app to ask for the {\tt SEND\_SMS} permission once the sending button is pushed, while an SMS message sent by a flashlight instance is suspicious.
Given a large number of popular apps with similar functionalities and user interfaces (UIs), 
{\tt COSMOS} is able to learn a generic model that reflects the correspondences between foreground user interface patterns (texts, layouts, etc.) and their background behaviors. 

However, such a one-size-fits-all model is not always sufficient. 
In practice, 
different users may have very different preferences on the same permission request even in a similar context~\cite{wagner2017,smarper}. 
Therefore, {\tt COSMOS} then incrementally trains the generic model on each device with its user's privacy decisions made over time. In the end, each user has a personalized model. 

In summary, this paper makes the following contributions: 
\begin{itemize}
\item We propose a novel permission system that inspects app foreground information to enforce runtime contextual integrity. Our approach involves a two-phase learning framework to build a personalized model for each user. 
\item We implement a prototype of the {\tt COSMOS} permission system. 
It is implemented as a standalone app and can be easily installed on Android devices with root access. 
It is also completely transparent to third-party apps. 
\item We show that {\tt COSMOS} achieves both high precision and high recall (95\%) for 6,560 requests from both authentic apps and malware.
Further, it is able to capture users' specific privacy preferences with an acceptable median f-measure (84.7\%) for 1,272 decisions collected from users.
We also show {\tt COSMOS} can be deployed on real devices to provide real-time protection with a low overhead.
\end{itemize}

\section{Problem Statement}\label{sec:prob}

\subsection{Threat Model}
We target threats from third-party apps that access {\it unnecessary} device resources 
in fulfilling their functionalities provided to the users. 
Such threats come from both intended malicious logic embedded in an app and vulnerable components of an app that can be exploited by attackers.
We assume that the underlying operating system is trustworthy and uncompromised.

\subsection{Design Goals}
Our goal is to design a runtime permission system that enforces \emph{contextual integrity} with \emph{minimum user involvement}.

\smallskip
\noindent \textbf{Contextual Integrity:}
To enforce contextual integrity in mobile platforms, one needs to ask the following three questions regarding a permission request:

\textit{Who initiated the request?}
An app may request the same permission for different purposes. For instance, a map app may ask user's locations for updating the map as well as for advertisement.
Although it can be difficult to know the exact purpose of a permission request, it is critical 
to distinguish the different purposes by tracing the sources of requests. 

\textit{When did it happen?}
Ideally, a permission should be requested only when it is needed, which implies that the temporal pattern of permission requests is an important piece of contextual data.
For instance, it is helpful to know if a permission is requested at the beginning or at the termination of the current app activity and if it is triggered by proper user interactions such as clicking, checking, etc.

\textit{What kind of environment?}
A proper understanding of the overall theme or scenario when a permission is requested is critical for proper permission control.
For instance, it is expected that different scenarios such as entertainment, navigation, or message composing may request very different permissions. In contrast to {\it who} and {\it when} that focus on detailed behavioral patterns, {\it what} focuses on a high level understanding of the context. 

The above context will help us detect mismatches between app behavior and user expectation.
A research challenge here is how to learn the context automatically for dynamic access control. Moreover, as user expectation may vary from one to another, how should we meet each user's personal expectation? 

\smallskip
\noindent \textbf{Minimum User Effort:}
Recent studies on runtime permission control focus on characterizing users' behavioral habit and attempt to mimic users' decisions whenever possible~\cite{smarper,wagner2015,wagner2017}.  Although this approach caters to an individual user's privacy preference, it also raises some concerns. First, a user could be less cautious and the potential poor decisions made by the user could lead to poor access control~\cite{wagner2017}.
Second, malicious resource accesses are user independent (although they may still be context dependent), which should be rejected by the runtime permission system without notifying the user.
Furthermore, the permission system should automatically grant the permissions required for the core functional logic indicated by the context of the running app to reduce user intervention.
To achieve minimum user involvement, our system should notify a user only when the decision is user dependent {\it and} the current scenario is new to the user. In all other cases, it should automatically accept or deny a permission request based on the current model with the user's previous decisions incorporated.


Our system should also provide high \textit{scalability} and {\it adaptivity}. It should scale to a large number of diverse permission requests and require no app source code or additional developer effort. Its accuracy and usability can be continuously improved 
with more user decisions incorporated. 

\begin{figure*}[htb]
\centering
\includegraphics[height=2.2in,width=0.9\textwidth]{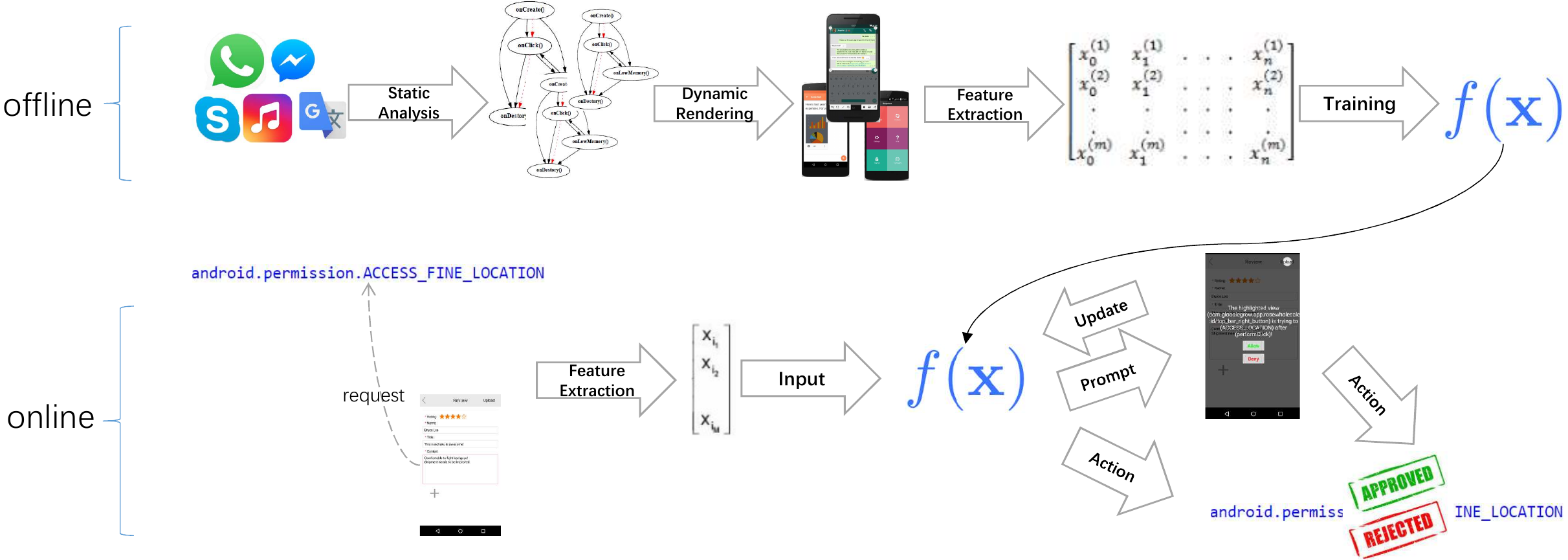}
\caption{System architecture}
\label{fig:arch}
\end{figure*}
\label{sec:overview}
\section{System Architecture}
Figure~\ref{fig:arch} depicts the overall system architecture of {\tt COSMOS}, which contains two phases. 

\smallskip
\noindent \textbf{Offline Phase}:
    The offline phase (details in Section~\ref{sec:offline}) builds a generic model to predicate user expectation when a sensitive permission request is made. To build the model, we collect a large number of benign apps and malicious apps and develop a lightweight static analysis technique to extract the set of sensitive API calls and the corresponding foreground windows. 
    Subsequently, the windows are dynamically rendered to extract their layouts as well as the information of their embedded widgets.
    The system calls, widgets and layouts are then used to extract features to build learning models that classify each sensitive API call of third-party apps as either legitimate, illegal or user-dependent.

\smallskip
\noindent \textbf{Online Phase}:
    In the online phase (discussed in Section~\ref{sec:online}), the generic model trained previously is personalized as follows. For each sensitive API call invoked by a third-party app, our mediation system will intercept the call and leverage the personalized model to identify its nature (initially, the personalized model is the same as the generic model). 
    The sensitive API call is allowed if it is classified as legal and is blocked (optionally with a pop-up warning window) if it is classified as illegal. 
    Otherwise, the API call is considered as undetermined and the user will be notified for decision making. The user's decision is then fed back to the online learning model so that automatic decisions can be made for similar scenarios in the future.
    To better assist user's decisions, detailed contextual information is provided in addition to the sensitive API call itself. Moreover, we provide specific mechanisms to handle background requests without foreground context. 

\ignore{
\subsection{Data Collection}
We perform the followings for the given app:
\begin{enumerate}
  \item Search the sensitive API calls.
  \item For each sensitive invocation, we retrieve the entry points and identify the Android components (e.g. an Activity) where the entries lie.
  We refer the extracted Activities as {\it the target Activities}.
  \item Instrument the app to make each target Activity callable from outside environment.
  \item Run each target Activity and record the UI information.
\end{enumerate}
}

\ignore{
\begin{minipage}{\textwidth}
  \begin{minipage}[htb]{0.6\textwidth}
    \centering
     \captionof{lstlisting}{
Code example}\label{lst:component}
\begin{lstlisting}
  public class ComposeView {
    public void onFinishInflate() {
        ...
        mButton = findViewById(compose_button);
        mButton.setOnClickListener(this);
    }

  public void onClick(View v) {
  		  if (mButtonState)
          	sendTextMessage(...);
    }
  }

  public class ComposeFragment {
    public View onCreateView( ...) {
      ...
      mComposeView.setLabel("Compose");
      ...
    }
  }
\end{lstlisting}
\end{minipage}
  \hfill
  \begin{minipage}[htb]{0.35\textwidth}
    \centering
    \includegraphics[height=2.9in]{example}
    \end{minipage}
  \end{minipage}
}

\begin{lstlisting}[caption=Code example, label=lst:component, float, floatplacement=H, basicstyle=\scriptsize]
  public class ComposeView {
    public void onFinishInflate() {
        ...
        mButton = findViewById(R.id.compose_button);
        mButton.setOnClickListener(this);
        ...
    }

  public void onClick(View v) {
          ...
          sendTextMessage(...);
          ...
    }
  }

  public class ComposeFragment {
    public View onCreateView( ...) {
      ...
      mComposeView.setLabel("Compose");
      ...
    }
  }
\end{lstlisting}

\section{Offline Analysis and Learning}\label{sec:offline}
This section discusses the process of building a generic permission model using program analysis and machine learning.
\subsection{Foreground Data Extraction}\label{sec:collection}

{\tt COSMOS} models the context of a sensitive request using the foreground data associated with the request. Although one can manually interact with an app and record the foreground data, 
it is infeasible to build a faithful model by analyzing a large number of apps manually. 
An alternative approach is using existing random fuzzing techniques
that generate random inputs in order to trigger as many sensitive behaviors as possible.
However, random fuzzing is inefficient, as it generates many inputs with similar program behavior.
More importantly, without any prior knowledge, random testing wastes time on exploiting code paths that are irrelevant to sensitive resource accesses. 

In this work, we propose a hybrid approach to collect relevant foreground data, including the set of widgets, the triggering events and the windows associated with sensitive API calls. Our approach has two phases, a \textbf{static analysis} phase and a \textbf{dynamic rendering} phase.
In particular, we adopt static program analysis to accurately locate the foreground components that would trigger a permission request.
Compared with random fuzzing, our approach achieves better coverage and eliminates redundant traces. 
The identified foreground components are then rendered dynamically with actual execution, which provides more complete and precise information compared to a pure static approach.
As an over-approximation approach, pure static analysis is criticized by generating false relationships between UI elements~\cite{borges2017data}.
Furthermore, we underline the fact that the existing hybrid approaches~\cite{leaksemantic} only focus on the program slices directly related to sensitive invocations, which typically omit the code corresponding to user interface.  

To illustrate our hybrid approach, we use the code in Listing~\ref{lst:component} as an example throughout this section, which presents the underlying logic of the open-source SMS app \texttt{QKSMS}. 

\smallskip
\noindent \textbf{Static Analysis:}
For each target app, we first identify its permission-protected API calls through method signatures.
We construct a call graph for the given app with the help of {\tt FlowDroid}~\cite{arzt2014flowdroid} and iterate over the graph to locate the target calls.
The list of permission-protected API methods is provided in {\tt PScout}~\cite{felt2011android} and {\tt FlowDroid}.
In {\tt QKSMS}, \texttt{sendTextMessage} in line 11 is marked as a sensitive API caller that requests the {\tt SEND\_SMS} permission.

The set of call graph entry points of the sensitive API calls are then identified by traversing through the call graph.
For instance, the \texttt{onClick} method inside {\tt ComposeView} (line 8) is found as an entry method of {\tt sendTextMessage}.

Further, the set of widgets that invoke the entry points (e.g, {\tt mButton}) are extracted by locating the event handlers of the entry points.
We then conduct a \emph{data flow analysis} to track the sources of the widgets.
After knowing where the widget {\tt mButton} is initialized, we are able to get its unique resource id ({\tt compose\_button}) within the app by inspecting the initialization procedure (line 4).

As the foreground windows set context, our analysis goes beyond individual widgets by further identifying the windows that the widgets belong to.
In our case, we aim to identify the {\tt Activity} that includes {\tt mButton}.
Since {\tt mButton} is initialized inside {\tt ComposeView}, we search for the usage of {\tt ComposeView} within the app.
{\tt ComposeView} is declared in {\tt ComposeFragment}, from which we can finally identify {\tt ComposeActivity} as the window for {\tt mButton}.

We notice that due to over-approximation, the static analysis phase may 
misidentify some UI elements that are not correlated with the indicated permission request.
We manually filter the misidentified samples before building the learning model to lower the impact of false alarms as much as possible.
However, we remark that it can be beneficial to keep some contextual instances that do not request a permission and label them as illegal since they simulate more scenarios that should not use the permission.

\smallskip
\noindent \textbf{Dynamic Rendering:}
For each target {\tt Activity} recognized by our static analysis (e.g., {\tt ComposeActivity}), we then render it with actual execution to precisely extract its layout and widget information.
Actual execution enables us to extract relevant data 
loaded at runtime.
Capturing rendering information specified by source code is intractable for static rendering approaches such as {\tt SUPOR}~\cite{supor}, which solely leverage app resource files to uncover the layout hierarchies.
For instance, the title of the crafting page ({\tt Compose}) of {\tt QKSMS}, a critical piece of context while using the app, is declared in the Java code (line 19 in Listing~\ref{lst:component}) instead of the resource files.
Losing this kind of dynamically generated information may hinder the progress of our upcoming task to precisely infer the purpose of the underlying program behavior.

Most {\tt Activities} cannot be directly called by default.
Hence, for each app, we automatically instrument the app configuration file {\tt manifest.xml} with a tag {\tt <android: exported>} and then repackage it into a new {\tt apk} file.
After installing the new package, we wake up the interested {\tt Activities} one by one with the {\tt adb} commands provided by Android.
Once an {\tt Activity} is awakened, the contextual foreground app data, including the layout and widget information, is extracted and stored in XML files.
For some {\tt Activities} that cannot be correctly started in this way, we can manually interact with them. 
Advanced automatic UI interaction is an active open research problem~\cite{borges2017data} and it goes beyond the scope of this paper. 


\subsection{Classification}\label{sec:learning}
\ignore{
\begin{table}[t]
\centering
\setlength\tabcolsep{0pt} 
\begin{threeparttable}
\caption{\label{tab:func} Legal Permission Usage}
\small{
\begin{tabular} {c c}
\toprule
Permission & Functionality  \\\midrule
{\tt LOCATION} & weather, map, navigation, tracking\\
        &    nearby services, location-based socialization \\
{\tt CAMERA} &  photo/video shooting, scanner, flashlight \\
        & video chatting, check deposit, face recognition \\
{\tt NFC} &  device binding, payment \\
{\tt RECORD\_AUDIO} & audio recording, speech recognition\\
        & audio chatting, sleep monitor, talking game \\
{\tt BLUETOOTH} & file transfer, accessory pairing\\
        & stream transmission\\
{\tt DEVICE\_ID} & identity verification, device configuration, tracking \\
{\tt SEND\_SMS} & chatting, text invitation, authentication\\
\bottomrule
\end{tabular}
}
\end{threeparttable}
\end{table}
}

Using the extracted foreground data, we are able to build a machine learning model to detect user-unintended 
resource accesses.
Given a permission request, we consider it as :

\smallskip
\noindent \textit{Legitimate:} if the permission is necessary to fulfill the core functionality indicated by the corresponding foreground context.
The requests in this category would be directly allowed by our runtime mediation system to eliminate unnecessary user intervention. We emphasize that the core functionality here is with respect to the running foreground context, not the app as a whole.
For example, some utility apps include a referral feature for inviting friends to try the apps by sending SMS messages. 
This is typically not a core functionality of the apps and the developers normally do not mention this feature on the apps' description pages.
However, the SMS messages sent under the ``invite friends'' page after a user clicks the {\tt Invite} button should be considered as user intended.
In contrast, description-based approaches~\cite{pandita2013whyper,gorla2014checking} would unnecessarily raise alarms.

\smallskip
\noindent \textit{Illegitimate:} if the permission neither serves the core functionality indicated by the foreground context nor provides any utility gain to the user.
An illegitimate request can be triggered by either malicious code snippet or flawed program logic.
The latter can happen as developers sometimes require needless permissions due to the misunderstanding of the official development documents~\cite{felt2011android}.

\smallskip
\noindent \textit{User-dependent:} if the request does not confidently fall into the above two categories; that is, it is not required by the core functionality suggested by the foreground context, but the user may obtain certain utility by allowing it.
Intuitively, in addition to the core functionality, the foreground context may also indicate several minor features that require sensitive permissions. Whether these additional features are desirable can be user dependent.
For example, besides the {\tt CAMERA} permission,
a picture shooting instance may also ask permissions such as {\tt ACCESS\_LOCATION} to add a geo-tag to photos. 
Although some users may be open to embed their location information into their photos that may be shared online later, those who are more sensitive to location privacy may consider this a bad practice.
In this case, we treat {\tt ACCESS\_LOCATION} as a user-dependent request and leave the decision to individual users. 

\smallskip
\noindent \textbf{Features:}
Before extracting features from the collected foreground contextual data,
we pre-process the crawled layouts to better retrieve their structural properties.
Mobile devices have various resolutions.
With absolute positions, models built for one device may not 
apply to other devices with different resolutions. 
Therefore, we divide a window into a $3 \times 3$ grid 
and map absolute positions to relative positions. 

The processed layouts are then used to extract features.
We construct three feature sets to enforce contextual integrity discussed in Section~\ref{sec:prob}.
More specifically, we derive the following features from a sensitive request: 

\smallskip
\noindent \textit{Who:}
The static phase of our foreground data collection described in Section~\ref{sec:collection} allows us to identify the widgets leading to sensitive API calls.
We then collect the attribute values of the target widgets using the dynamically extracted layout files.
It is possible that the permission request is triggered by an {\tt Activity} rather than a widget.
In this case, we would leave the value of this feature set as empty and rely on the ``what'' feature set to handle windows.

\smallskip
\noindent \textit{When:}
The call graph traversal gives us entries of sensitive API calls. An entry point can be either a lifecycle callback or an event listener. The lifecycle models the transition between states such as the creation, pause, resume and termination of an app component. 
The event listeners monitor and respond to runtime events.
Both lifecycle callbacks and event listeners are prior events happened before an API call and serve as useful temporal context to the call. We therefore use the signatures of entry methods as the ``when'' feature set.

\smallskip
\noindent \textit{What:}
The text shown on target widgets could be too generic (such as {\tt Ok} and {\tt Yes}) to convey any meaningful context.
Therefore, we also derive features from the windows to help infer the overall theme of the requesting environment.
We iterate over the view hierarchy of the window layout and obtain all the related widgets with text labels. For every such widget, 
we save the text on the widget and its relative position in the window as features.
Including both textual and structural attributes 
provides better scalability to capture semantic and structural similarities across millions of pages.
Although developers may adopt various design styles for the same functionality, their implementations usually share a similar characterization.
For instance, we do not need to know whether a window is implemented with {\tt Material} design. Instead, learning the title shown at the top of the window, such as {\tt Compose} and {\tt New message}, is crucial.


By focusing on features directly visible to users, our approach is resilient to code level obfuscation. 
Note that the entry methods are overridden of the existing official SDK APIs and cannot be renamed by the third parties.

For each of the three feature sets mentioned above, we generate a separate feature vector. 
Note that although attributes of a widget leading to sensitive API calls appear in both the ``who'' feature set and the ``what'' feature set, 
they are treated separately to stress the triggering widget.
For the ``what'' set, the text and positions of all the widgets shown on the window are included, while for the ``who'' set, only those related to the triggering widget are included.
All the textual features are pre-processed using natural language processing (NLP) techniques. 
In particular, we perform {\it identifier splitting, stop-word filtering, stemming} and leverage bag-of-words model to convert them into feature vectors. The process is similar to other text-based learning methods~\cite{flowintent}.
In addition to the three sets of features, it is possible to include more features to further raise the bar of potential attacks.

\ignore{
\begin{table}[]
\centering
\setlength\tabcolsep{0.5pt}
\begin{threeparttable}
\caption{\label{tab:features}The Feature Sets}
\small{
\begin{tabular}{ l | c | r }
\toprule
Type & Source & Features  \\\midrule
{\tt What} & Window & textual surroundings\\
& & layout  \\
{\tt When} & Activation Event & event type\\ 
{\tt Who} & Sensitive Widget & attribute, position \\
{\tt Namespace} & Sensitive API call and caller &  class name \\
 & & method name\\
\bottomrule
\end{tabular}
}
\end{threeparttable}
\end{table}
}

\smallskip
\noindent \textbf{Training:}
\label{sec:learning}
Using the three sets of features discussed above, we train a one-size-fits-all learning model as follows. 
For each permission type, a classifier is trained with a data mining tool {\tt Weka}~\cite{weka} using the manually labeled sensitive API calls related to that permission.
The classifiers are trained separately for different permissions to eliminate potential interference. Each permission request is labeled as either \emph{legal} or \emph{illegal} based on the foreground contextual data we collected including: the entry point method signature, the screenshot of the window, and the highlighted widget invoking the API call (if there is such a widget). We ensure contextual integrity by checking whether they altogether imply the sensitive API call. The request is marked as illegal if it is not supported by any type of the foreground data. For instance, a {\tt SEND\_SMS} permission requested under the ``Compose'' page without user interactions or required by an advertisement view is categorized as illegal.

As we mentioned in Section~\ref{sec:overview}, our generic models will be continuously updated at runtime to incorporate individual user's preferences.
One option is to keep sending data to a remote cloud for pruning the models. However, since the content shown on a device can be deeply personal, transmitting this kind of sensitive data out of the device would raise serious concerns on potential leaks~\cite{fernandes2016appstract}.
Consider the SMS composing example again, the window may contain private information typed by the user, which is inappropriate to share with a third-party service.
On the other hand, the limited computational power of mobile devices makes it infeasible to repeatedly train complicated models from scratch inside the devices.
To meet both the privacy and performance requirements, we apply light-weight incremental classifiers that can be updated instantaneously using new instances with a low 
overhead, which matches the memory and computing constraints of smart phones~\cite{yin2016incremental}.
A key question is which incremental learning technique to use.
To this end, we have evaluated popular incremental learning algorithms. The detailed results are given in Section~\ref{sec:eval}.

\ignore{
\begin{figure}[htb]
\centering
\includegraphics[height=2.0in,width=0.48\textwidth]{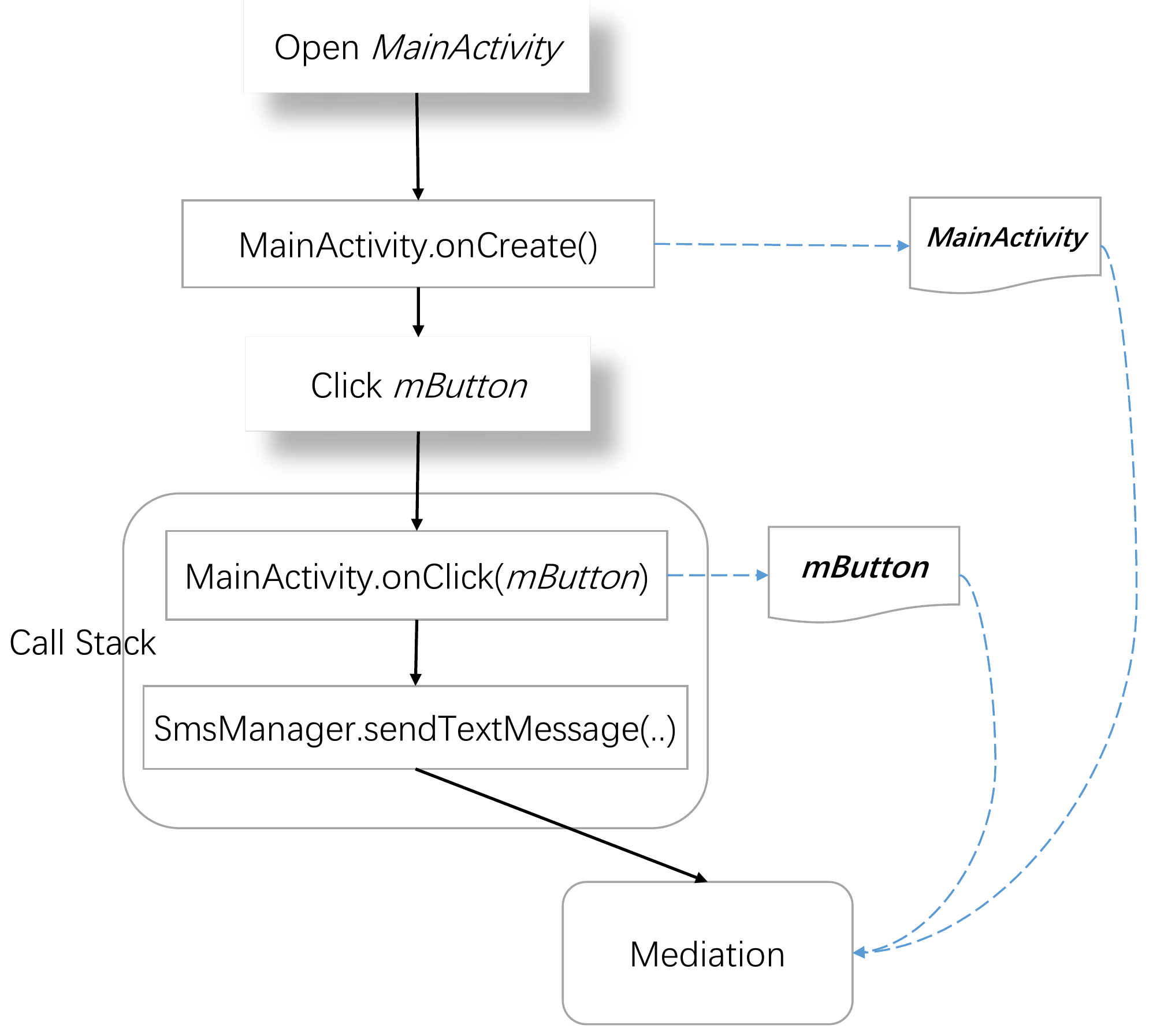}
\caption{Online extraction}
\label{fig:online}
\end{figure}
}

  \begin{figure*}[htb]
  \begin{minipage}[b]{0.45\textwidth}
    \centering
    \includegraphics[height=2.0in]{fig_online.pdf}
    \captionof{figure}{Online extraction}\label{fig:online}
  \end{minipage}
  \hfill
  \begin{minipage}[b]{0.55\textwidth}
    \centering
    \includegraphics[height=1.9in]{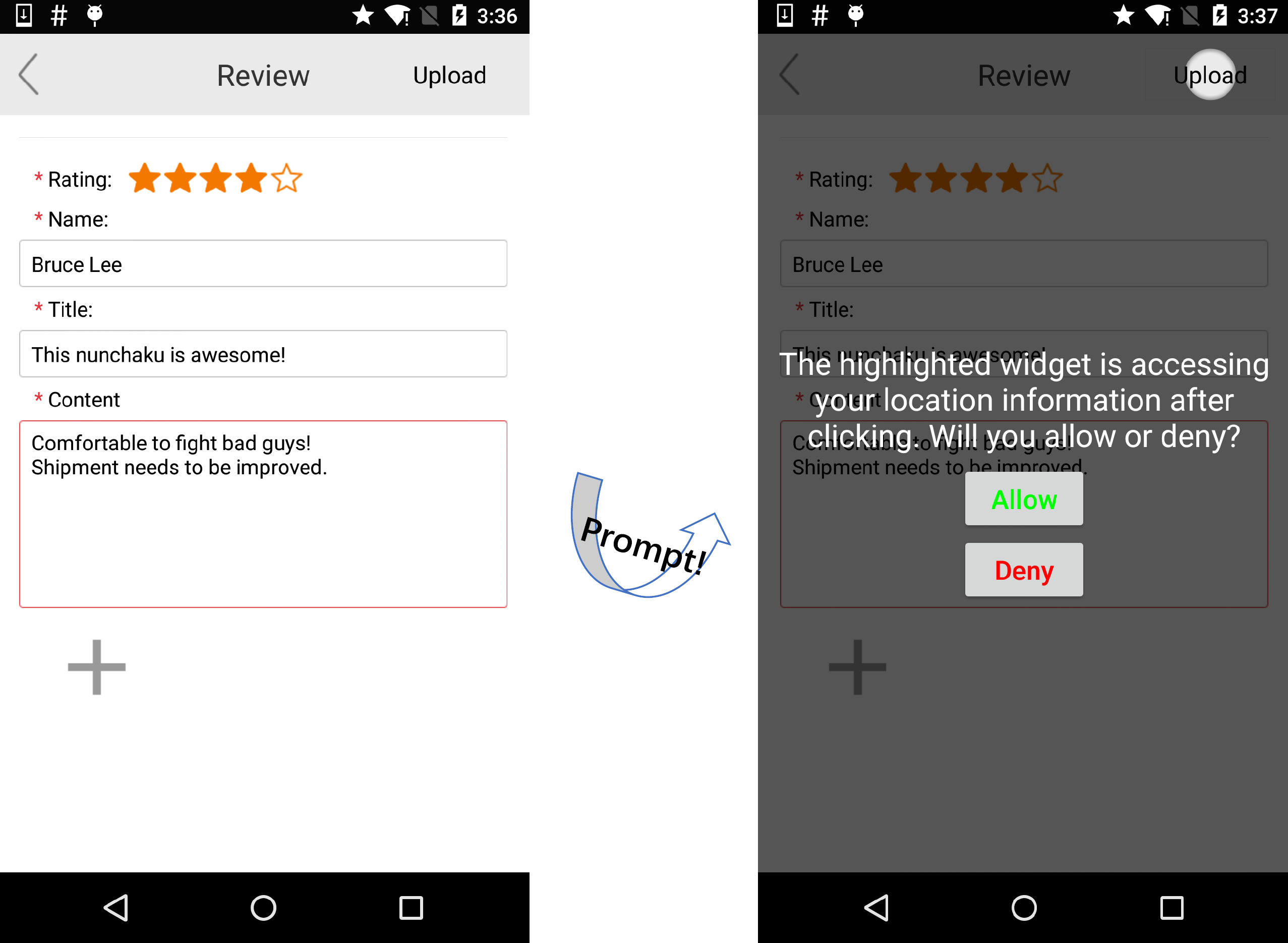}
      \captionof{figure}{An example prompt shown by {\tt COSMOS}. In the top right corner, the {\tt Upload} button that is accessing the location is highlighted.}\label{fig:prompt}
    \end{minipage}
  \end{figure*}

\section{Online Permission System} \label{sec:online}
{\tt COSMOS} as a mediation system dynamically intercepts sensitive calls, collects features for them, and finally automatically grants or denies the requests using online learning models.


\subsection{Mediation and Data Extraction}
Android does not officially allow a third-party app to mediate other apps' requests.
Instead of modifying the OS and flashing the new firmware, {\tt COSMOS} is written in Java as a standalone Android app and can be easily installed on Android devices with root access.
The implementation of {\tt COSMOS} is based on {\tt Xposed}~\cite{xposed}, an open-source method hooking framework for Android. {\tt Xposed} provides native support to intercept method calls, which enables us to execute our code before and after execution of the hooked method.

To detect improper permission requests at runtime, {\tt COSMOS} dynamically extracts information from the UI elements associated with sensitive calls.
Consider the example shown in Figure~\ref{fig:online}. The {\tt sendTextMessage} is triggered after clicking {\tt mButton} shown on {\tt MainActivity}.
{\tt COSMOS} needs to retrieve the memory references of the interested UI elements, including the running instances of {\tt mButton} and {\tt MainActivity}.
However, simply intercepting the target sensitive call is insufficient.
The problem is that although we can extract the values of the variables appeared in the current call (e.g., {\tt sendTextMessage}), retrieving the values from the prior calls (e.g., {\tt onClick}) is currently infeasible in {\tt Xposed}, which makes it difficult to retrieve the trigger UI instances by only hooking the sensitive API call.

To address the above problem, {\tt COSMOS} intercepts the invocations of both Activity lifecycle callbacks (e.g., {\tt Activity.onCreate}) and event listeners (e.g., {\tt onClick}) in addition to sensitive API calls. For each of these methods, it records the references of the method parameters. 
For instance, in the above example, the references to {\tt mButton} and the Activity are stored when processing {\tt onClick(mButton)}.
When it encounters a sensitive API call, {\tt COSMOS} retrieves the {\it latest} widget and Activity it saved, and extracts the same features from them as in the offline model. In particular, ``who'' features are collected from the widget and ``what'' features are extracted from the activity by iterating over all its widgets.
Moreover, {\tt COSMOS} examines call stack traces to determine the entry point methods leading to sensitive calls, which are used to derive the ``when'' features.

After converting the features into numerical values, {\tt COSMOS} uses the online learning model to predict the type of the sensitive request. 
It automatically grants the permission if the request is classified to be legitimate with high confidence
and rejects the request if it is confidently classified as illegitimate.
For a rejected request, {\tt COSMOS} further pops up a warning to the user including the details of the request.
A request that is neither legal or illegal with high confidence will be treated as user-dependent and will be handled by the user preference module as discussed below.

As users can switch between Activities, a request may be initiated by a background Activity. By tracking the memory references of the associated UI elements, \emph{{\tt COSMOS} is able to reason about the background requests even if the associated UI elements are currently invisible.}


\ignore{
\begin{figure}[t]
\centering
\includegraphics[height=1.9in,width=0.53\textwidth]{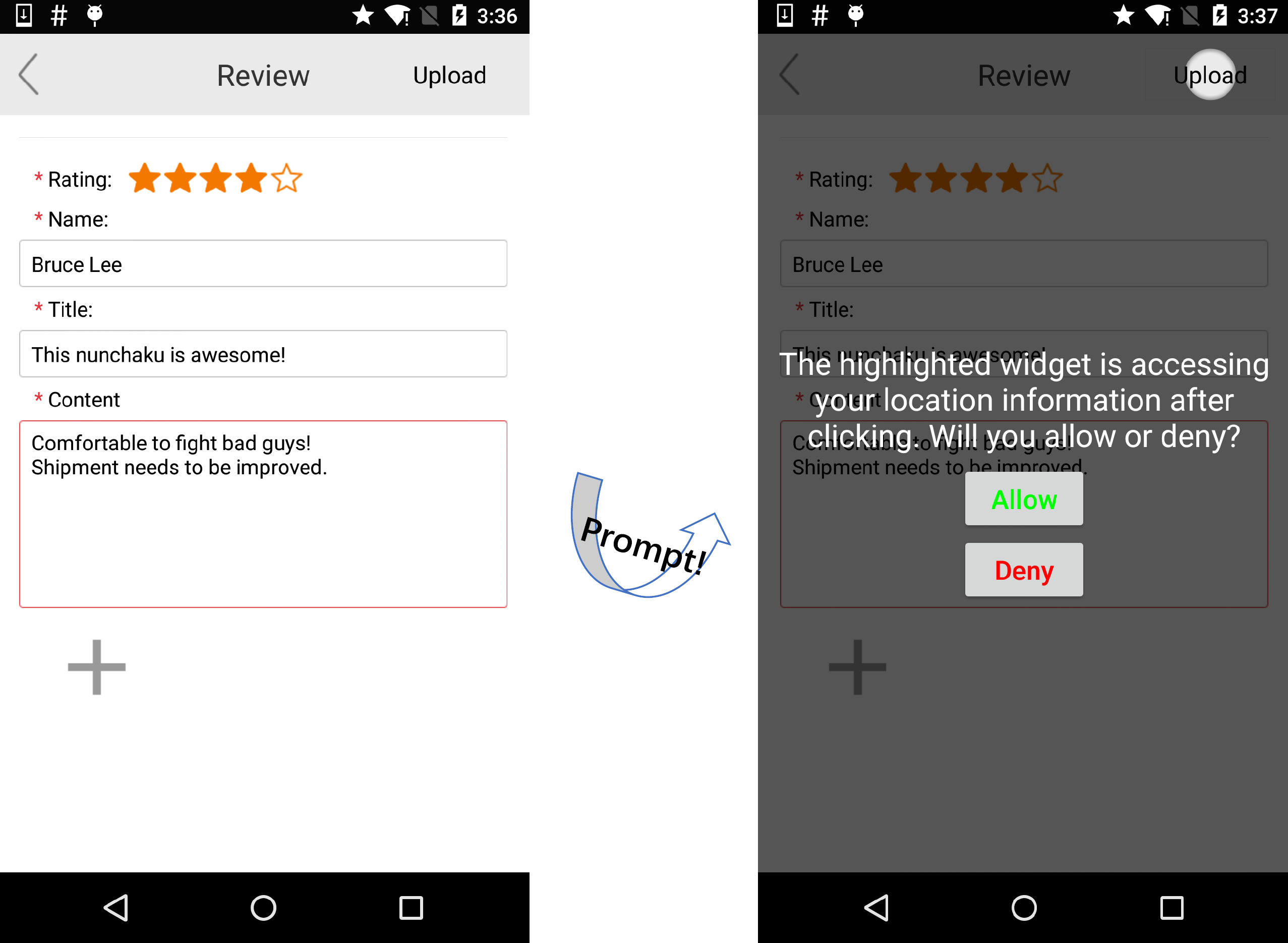}
\caption{An example user prompt shown by {\tt COSMOS}. In the top right corner, the ``Upload'' button that is accessing the device location is highlighted.}
\label{fig:prompt}
\end{figure}
}
\smallskip
\noindent \textbf{GUI Spoofing:}
To ensure that the foreground data is indeed associated with the background request, {\tt COSMOS} dynamically inspects the widget information with the hook support and ignores the widgets that are not owned by the permission requesting app.
Thus, {\tt COSMOS} is resilient to GUI spoofing that tries to evade detection by hiding behind the interfaces of other apps.

More advanced GUI spoofing attacks have also been proposed in the literature~\cite{bianchi2015app}.
For example, when a benign app running in the foreground expects a sensitive permission to be granted, a malware may replicate and replace the window of the benign app to elicit the user.
However, such attacks can be hard to implement in practice as they require Accessibility feature enabled to the malware by the user.
It is worth noting that using Accessibility may play against the malware itself, since Android repeatedly warns the user about the threats caused by Accessibility.
If needed, {\tt COSMOS} can also intercept the calls initiated from Accessibility to further alarm users.

\smallskip
\noindent \textbf{Background Services:}
An Activity can start a background Service. 
However, when a sensitive call is initiated by a  Service, its call stack does not contain the information of the starting Activity.
In this case, {\tt COSMOS} monitors the calls of {\tt Activity.startService(Intent)} to track the relationship between running Activities and Services.
{\tt COSMOS} then uses the information available from the Activity to infer the purpose of a Service request.

A Service may exist without any triggering Activity. In this case, {\tt COSMOS} notifies the user about the background request and lets the user decide whether to allow or deny the request.
Alternatively, we can always reject such requests.
We argue that sensitive services should not exist unless they provide sufficient foreground clues to indicate their purposes.
Users tend to reject requests without foreground as suggested by three recent user studies~\cite{wagner2015,wagner2017,smarper}. Indeed, 
the recent updates of Android further restrict background services~\cite{androido}.

\subsection{User Preference Modeling}

To incorporate user preferences, {\tt COSMOS} notifies the user if the online model identifies a request as user-dependent. Consider the example shown in Figure~\ref{fig:prompt}.
The UI shows a product review page and a location permission is requested once the {\tt Upload} button is clicked. On the one hand, the user may be beneficial from sharing location if the seller provides subsequent services to promote customer experience based on the user's review and location. On the other hand, the sharing behavior could put the user at risk since there is no guarantee how exactly the location information would be used by the app developer.
As the page does not provide enough evidences whether location sharing is necessary, {\tt COSMOS} treats the instance as user-dependent, and then creates a prompt to accept user decision. Our prompt not only alarms the user about the existence of the permission request, but also highlights the widget that triggered the request and the activation event. 



The user decision, along with the features of the instance, is then used to update our model.
Discussed in Section~\ref{sec:learning}, our classifiers are built through incremental learning in order to take care of both privacy concern and performance overhead.
The incremental learning model immediately accepts the new instance and adjusts the decision strategy to better match user criteria next time.
\begin{table*}
\begin{minipage}[b]{0.5\linewidth}
\centering
\small{
\captionof{table}{Results for Different Classifiers}\label{tab:overall}
\begin{tabular}{ l c c r }
\toprule
Algorithm &  \tabincell{c}{Median \\ F-measure} & \tabincell{c}{Average \\ Precision} & \tabincell{c}{Average \\ Recall}  \\\midrule
HT & 77.9\% & 81.7\%  & 78.3\%  \\
NB & 93.9\% &  93.3\%  & 92.9\% \\
SVM & 95.5\% & 95.4\% & 95.4\% \\
LR & 96.1\% & 95.8\% & 95.5\% \\
\bottomrule
\end{tabular}
}
\end{minipage}
\begin{minipage}[b]{0.5\linewidth}
\centering
\small{
\captionof{table}{Results for Different Permissions}\label{tab:perm}
\begin{tabular}{ l c c  r }
\toprule
Permission & Precision & Recall & F-Measure  \\\midrule
{\tt DEVICE\_ID} & 89.8\%   &  89.3\% & 89.3\%  \\
{\tt LOCATION} & 93.8\%  &  93.9\% &   93.8\% \\
{\tt CAMERA} & 95.0\%   &   95.0\% &   95.0\% \\
{\tt RECORD\_AUDIO} &  96.0\%  & 96.1\%  & 96.1\%  \\
{\tt BLUETOOTH} & 97.9\% & 97.9\% & 97.9\%   \\
{\tt NFC} & 96.7\%  & 96.6\%  &  96.6\% \\
{\tt SEND\_SMS}  & 99.8\%   &  99.8\%  &  99.8\%  \\
\bottomrule
\end{tabular}
}
\end{minipage}
\end{table*}

\ignore{
 \begin{figure}
    \centering
        \includegraphics[height=1.8in]{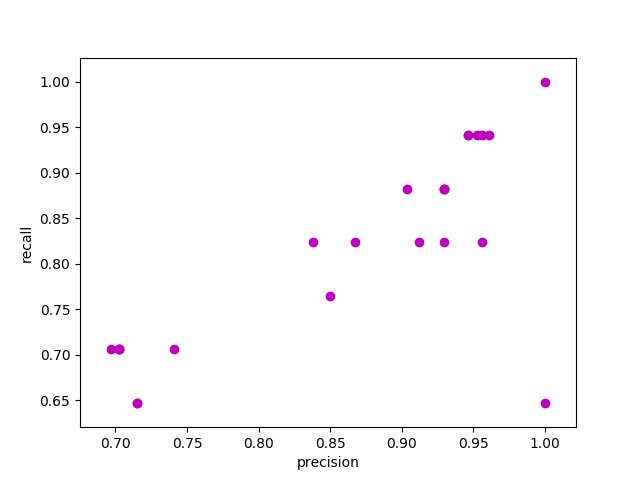}
      \captionof{figure}{The precision and recall of each participant. }\label{fig:user}
   \end{figure}
 
\begin{table*}
\begin{floatrow}
 \captionof{table}{Classification with Different Feature Sets}\label{tab:feature}
\small{
\begin{tabular}{ l  c c  r }
\toprule
Feature Type &  Precision & Recall & F-Measure  \\\midrule
\bf{Who} & 81.9\% &     78.8\%  &  75.7\% \\
\bf{When} &  69.7\%   &   70.7\% &   70.0\%  \\
\bf{What} & 95.4\%   &   95.3\% &   95.3\% \\
\bf{Who \& When}  & 80.0\%  &    79.1\%  &  76.9\%  \\
\bf{Who \& What} &   95.6\%   &   95.6\% &   95.6\%   \\
\bf{When \& What} &   95.6\%   &   95.6\% &   95.6\% \\
\bf{All} &   96.0\%  &    96.1\%  &  96.1\% \\
\bottomrule
\end{tabular}}
\end{floatrow}
\end{table*}
}
   
\begin{figure*}
\begin{minipage}{0.5\linewidth}
\centering
 \includegraphics[height=1.8in]{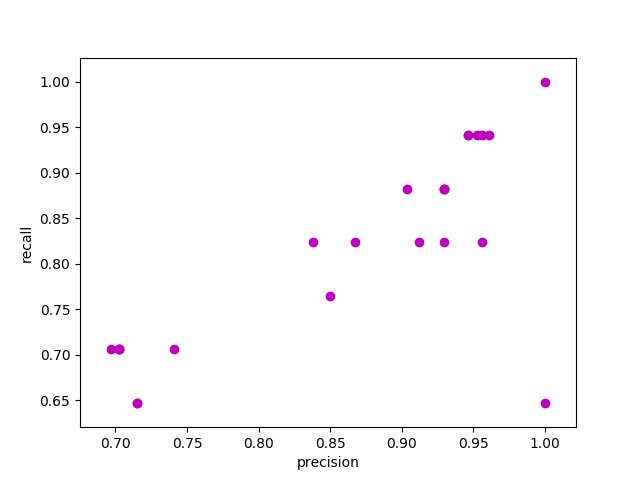}
      \captionof{figure}{The precision and recall of each participant.}\label{fig:user}
\end{minipage}
\begin{minipage}{0.5\linewidth}
\centering
 
\small{
\captionof{table}{Classification with Different Feature Sets}\label{tab:feature}
\begin{tabular}{ l  c c  r }
\toprule
Feature Type &  Precision & Recall & F-Measure  \\\midrule
\bf{Who} & 81.9\% &     78.8\%  &  75.7\% \\
\bf{When} &  69.7\%   &   70.7\% &   70.0\%  \\
\bf{What} & 95.4\%   &   95.3\% &   95.3\% \\
\bf{Who \& When}  & 80.0\%  &    79.1\%  &  76.9\%  \\
\bf{Who \& What} &   95.6\%   &   95.6\% &   95.6\%   \\
\bf{When \& What} &   95.6\%   &   95.6\% &   95.6\% \\
\bf{All} &   96.0\%  &    96.1\%  &  96.1\% \\
\bottomrule
\end{tabular}
}
\end{minipage}
\end{figure*}
\section{Evaluation}\label{sec:eval}
In this section, we evaluate
the effectiveness of {\tt COSMOS} by answering the following questions:

  \noindent \textbf{RQ1}: Can {\tt COSMOS} effectively identify misbehaviors (i.e., inconsistencies between context and request) in mobile apps? How do the feature sets of \emph{who}, \emph{when} and \emph{what} contribute to the effectiveness of misbehavior identification?

  \noindent \textbf{RQ2}: Can {\tt COSMOS} be applied to capture personal privacy preferences?  
  
  \noindent \textbf{RQ3}: Can {\tt COSMOS} be deployed on real devices with a low overhead?

RQ1 measures the effectiveness of the generic models where individual user preferences are not involved.
A request that cannot be confidently labeled as either legal or illegal is considered as user-dependent and its relavant effectiveness is measured in RQ2. 


\subsection{RQ1: Accuracy in Identifying Misbehaviors}
We manually labeled 6,560 identified permission requests that belong to 1,844 different apps, each of them was either a top-ranked app crawled across 25 categories from Google Play, or a malware sample collected from VirusShare~\cite{virusshare}.
Each request was labeled through the associated foreground contextual data, including the widget (if any), the events and the window. 
In particular, we determined whether a request (e.g., {\tt RECORD\_AUDIO}) was initiated by an appropriate widget (e.g., a ``microphone'' button) after a proper interaction (e.g., clicking) and under a correct environment (e.g., voice assistant).


\smallskip
\noindent \textbf{Overall Effectiveness:}
For each permission type, we leveraged the labeled requests both as training and test data in a five-fold cross validation.
Specifically, we randomly divided all instances of the same permission into 5 equally sized buckets, training on 4 of the buckets, and using the remaining bucket for testing.
We repeated the process 5 times and every bucket was used exactly once as the testing data.

As our online learning approach is a continuous training process that adapts to user decisions, a classifier that can process one example at a time is desired.
To determine which machine learning technique to use, we evaluated the effectiveness of four commonly used learning methods that support incremental classification, including
{\it Hoeffding Tree, (Multinomial) Naive Bayes, (linear) SVM} and {\it Logistic Regression}.
Compared to non-updatable classifiers, all these methods can iteratively incorporate new user feedback to update their knowledge and do not assume the availability of a sufficiently large training set before the learning process can start~\cite{ross2008incremental}.

A summary of the results is given in Table~\ref{tab:overall}, where the mean values are calculated over all permission types.
As we can see, logistic regression achieved the best result among all four classifiers.
Table~\ref{tab:perm} further provides detailed results of logistic regression on each permission type.
We considered seven permissions that are highly security or privacy sensitive~\cite{smarper,arzt2014flowdroid} and are commonly required by the collected apps.
We observed that among all the permission types, differentiating requests of {\tt DEVICE\_ID} is more challenging since developers normally do not provide sufficient information in apps to indicate why the permission is requested.
More human intervention could be beneficial regarding {\tt DEVICE\_ID}. 

\smallskip
\noindent \textbf{Feature Comparison:}
To measure how each feature set contributes to the effectiveness of behavior classification, 
we used the same learning technique (e.g., logistic regression) with different feature sets under  ``who'', ``when'' and ``what'' and some combinations of them, respectively.
The cross validation results of {\tt RECORD\_AUDIO} are presented in Table~\ref{tab:feature}.
Since the comparison results of other permissions share the similar trend, we omit them here.

For each feature set, we evaluated its effectiveness by comparing the evaluation metrics of our learning models when the feature set is used and when it is not. 
We found that the ``what'' features contributed the most among the three feature sets.
As we mentioned in Section~\ref{sec:intro}, benign instances often share similar themes that can be inferred from window content and layout.
For example, an audio recorder instance typically has a title {\tt Recorder}, a timer frame {\tt 00:00} at the center and two buttons with words {\tt start} and {\tt stop}, respectively. 
From these keywords and their positions in the page, {\tt COSMOS} is often able to tell whether the user is under a recording theme.
Although the ``what'' features successfully predicted most audio recorder instances, it may be of limited use in other cases where {\tt RECORD\_AUDIO} permission is used. 
For instances, developers tend to integrate voice search into their apps to better serve users.
However, as the searching scenarios differ greatly from each other, it is hard to classify their intentions using ``what'' features only.  

The ``who'' features help alleviate the above problem by further examining the meta data of the corresponding widget. 
For instance, {\tt co.uk.samsnyder.pa: id/speakButton} is an image button for speech recognition, which does not provide useful ``what'' features as the image button does not contain any extractable textual information.
However, the word ``speak'' in the resource-id clearly indicates the purpose of the button. In addition to the textual data, the relative position and the class attribute of a widget can also help locate non-functional components, e.g., the advertisements at the bottom.

%
We observed that for {\tt RECORD\_AUDIO}, the ``who'' features and the ``when'' features are highly correlated in most cases. This is because most sensitive method calls initiated by widgets are bound with the event {\tt onClick}. 
However, there are exceptions.
For instance, 
a walkie talkie app that transfers users' audio information to each other has the tips {\tt Press \& Hold} shown in its main window, which indicates that the recording should start only after user clicking. However, it actually starts recording once the app is open. This misbehavior can be effectively identified using the ``when'' features, which emphasizes that apps should request a permission only after proper user interactions.

\ignore{
Code obfuscation and name manipulation threaten the feasibility of the methods mainly rely on code-level namespace.
Shown in the Table~\ref{tab:feature}, the effectiveness of namespace can be highly improved with the combination of the features from user-interface.
From another perspective, namespace also contributes to the effectiveness of user-interface features.
The class name and method name of the sensitive API methods sometimes contain rich information to further eliminate ambiguous.
Recall the button {\tt co.uk.samsnyder.pa:id/speakButton} used for voice recognition, it would be more innocent if it calls
{\tt <android.speech.SpeechRecognizer:} {\tt setRecognitionListener(...)>}, instead of {\tt <android.media.AudioRecord:} {\tt startRecording()>}.
Note that an adversary could only manipulate the names of its self-defined API callers, not the API methods set by official SDK.
}

In summary, ``what'' features work well in differentiating between most legitimate and illegitimate instances at the current stage.
However, as malware continues to evolve, we expect that collecting more comprehensive contextual data including ``who'', ``when'' and ``what'' can provide better protection. 
The last row in Table~\ref{tab:feature} shows that the combination of all the three feature sets provides the best results. 

\begin{table}
\centering
\begin{threeparttable}
\caption{\label{tab:thread} }
\begin{tabular}{ l c c }
\toprule
Target App & Requests/min & \tabincell{c}{CPU Time (\%)} \\\midrule
\tabincell{c}{Wechat} & 12.6 & 4.4\% \\
\tabincell{c}{Yelp} & 5.8 & 2.2\% \\
Yahoo Weather & 2.5 & 1.4\% \\
\tabincell{c}{Amazon} & 0.8 & 0.6\% \\
\tabincell{c}{Paypal} & 0.4 & 0.2\% \\
\bottomrule
\end{tabular}
\end{threeparttable}
\end{table}

\subsection{RQ2: Effectiveness of Capturing Personal Preferences}
\ignore{
\begin{figure}
\centering
\includegraphics[height=2.2in,width=0.48\textwidth]{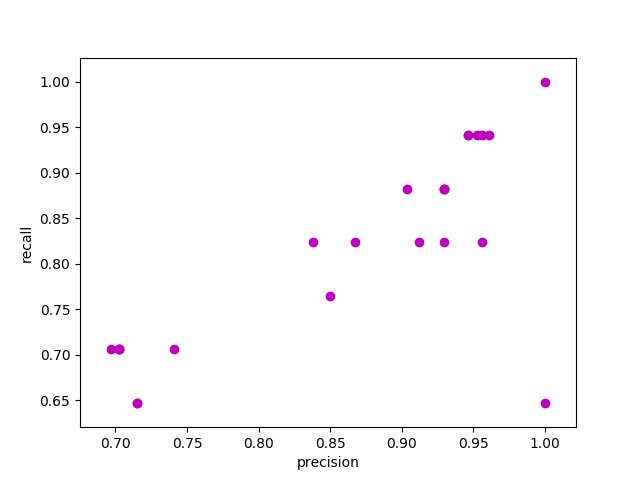}
\caption{The precision and recall of each user. We observed that some results were close and even identical, leading to the overlapping dots shown on the diagram. }
\label{fig:user}
\end{figure}
}

We conducted a lab-based survey \footnote{Our user study was proceeded with Institutional Review Board (IRB) approval.} to measure the effectiveness of our models to capture individual user's preferences, where we asked participants to classify a set of requests that were not faithfully labeled as legal or illegal. 
The survey was composed and spread through Google Forms.   
Among the 24 participants, 3 were professors, 6 were undergraduate students and 15 were graduate students.
Each user is asked to classify 50 location accessing requests collected from 40 real apps, covering several user-dependent scenarios such as shopping, photo geo-tagging, news, personal assistant and product rating. We collected 1,272 user decisions in total.

To simulate the real decision making on device, for each request, the following information is displayed to the participants:
1) Screenshot: the screenshot taken from the app right after the request was initiated, with the triggering widget highlighted.
2) Prior event: the event led to the request, such as app start and user clicking.
3) Meta-information: the app name and a Google Play link are included, whereby the participants can find more information. 


We evaluated the effectiveness of our user preference modeling by updating the pre-trained model constructed during the evaluation phase of RQ1 with the decisions collected from each individual user.
For each user's decisions, we randomly partitioned them into three sets and used two of the three sets as the training set to update the pre-trained model, and the rest set as the testing set.
Our model yielded a median f-measure of 84.7\% among the 24 users, which is reasonably good due to the limited number of samples.
We expect our model to be more accurate with more user feedback in the future.

Figure~\ref{fig:user} presents the detailed result of each individual.
We observed that some results were close and even identical, leading to the overlapping dots shown on the diagram.
A quarter of users' results have more than 90\% precision and 90\% recall.
Our model performed surprisingly well for one individual, with 100\% precision and 100\% recall. 
One individual tends to behave conservatively by rejecting nearly all requests, giving a sharp outlier in the lower right corner with a perfect precision but a terrible recall.
We also observed that some users made inconsistent decisions under a similar context.
For instance, one user allowed a request from a product rating page but rejected another with a closely related context. 
One possible explanation is that sometimes users are less cautious and make random decisions as suggested in~\cite{wagner2017}. Fortunately, our system can greatly help protect users from malicious behaviors caused by malware even if users make random decisions, since our generic model has already learned many misbehaviors in offline training. 

We also conducted a controlled experiment to test whether the fine-grained contextual info shown in our prompts can help users make better decisions.
We used the screenshots with location-based functionality at the center and a behavioral advertisement
at the bottom. Without prompts, 79.2\% of the participants chose to grant the permission. After being alerted that the location requests were actually initiated by advertisements, 73.9\% of them changed their minds to reject the requests.
These results encourage the deployment of {\tt COSMOS} to better assist users against unintended requests.



\subsection{RQ3: Performance Measurements}
To investigate the performance overhead incurred by {\tt COSMOS}, we installed five selected representative apps collected from different categories, including {\tt Wechat, Yelp, Yahoo Weather, Amazon} and {\tt Paypal}, on Nexus 5 with {\tt COSMOS} deployed.
We then interacted with them as in common daily use, and monitored the overhead introduced by {\tt COSMOS}.
Shown in Table~\ref{tab:thread}, {\tt COSMOS} consumed 1.8\% total CPU time on average and less than 5\% total CPU time for all the monitored apps. 
We also measured other impacts related to the performance such as memory and storage overhead, and all the values were reasonably small for daily use.
Details are omitted due to the page limit.

\section{Related Work}
Early studies on building context-aware systems mainly depend on manually crafted policies specific to certain behaviors~\cite{peg,miettinen2014conxsense,zhang2016rethinking}.
Recent approaches attempt to infer context-aware policies from users' behavioral traits~\cite{wagner2015,wagner2017,smarper}.
They observe that the visibility of apps is the most crucial factor that contributes to users' decisions on permission control. However, they do not capture more fine-grained foreground information beyond visibility and package names.

Some recent efforts have also been made to detect unexpected app behavior from UI data. For instance, {\tt AppIntent}~\cite{appintent} uses symbolic execution to extract a sequence of GUI manipulations leading to data transmissions. {\tt PERUIM}~\cite{peruim} relates user interface with permission requests through program analysis.  
Both approaches require user efforts to locate suspicious behaviors. 
{\tt AsDroid}~\cite{asdroid} identifies the mismatch between UI and program behavior with heuristic rules.
{\tt DroidJust}~\cite{chen2015droidjust} tracks the sensitive data flows to see whether they are eventually consumed by any human sensible API calls.
Ringer et al.~\cite{ringeraudacious} design a GUI library for Android to regulate resource access initiated by UI elements.
As these approaches rely on a small set of human crafted policies, they can only recognize certain misbehaviors within the domains.

Most recently, machine learning has been used to automate the analysis of user interface.
{\tt FlowIntent}~\cite{flowintent} and {\tt Backstage}~\cite{backstage} detect behavioral anomalies by examining all textual information shown on the foreground windows with supervised learning and unsupervised learning, respectively.
Though similar in spirit, they touch upon a subset of the challenges that {\tt COSMOS} tries to address and only focus on static app auditing. 
We extend this line of research in several ways.
First, we propose to protect contextual integrity through analyzing UI data from three distinctive perspectives: {\it who}, {\it when} and {\it what}.
Second, we provide a two-layer machine learning framework that can automatically grant the necessary permission requests and reject the harmful requests without requiring user involvement, as well as improving the decision accuracy based on user feedback.
Third, we implement our system on real devices to provide runtime protection and conduct comprehensive evaluations. 

\section{Conclusion}\label{sec:conclusion}
We propose a context-sensitive permission system called {\tt COSMOS} that automatically detects semantic mismatches between foreground interface and background behavior of running mobile applications. 
Our evaluation shows that {\tt COSMOS} can effectively detect malicious resource accesses with high precision and high recall. 
We further show that {\tt COSMOS} is capable of capturing users' specific privacy preferences and can be installed on Android devices to provide real-time protection with a very low performance overhead.

\section*{Acknowledgment}
The effort described in this article was partially sponsored by the U.S. Army Research Laboratory Cyber Security Collaborative Research Alliance under Contract Number W911NF-13-2-0045.
The views and conclusions contained in this document are those of the authors, and should not be interpreted as representing the official policies, either expressed or implied, of the Army Research Laboratory or the U.S. Government.
The U.S. Government is authorized to reproduce and distribute reprints for Government purposes, notwithstanding any copyright notation hereon.
The work of Zhu was supported through NSF CNS-1618684.

\footnotesize
\bibliographystyle{abbrv}
\bibliography{ref}

\begin{thebibliography}{10}

\bibitem{arzt2014flowdroid}
S.~Arzt, S.~Rasthofer, C.~Fritz, E.~Bodden, A.~Bartel, J.~Klein, Y.~Le~Traon,
  D.~Octeau, and P.~McDaniel.
\newblock Flowdroid: Precise context, flow, field, object-sensitive and
  lifecycle-aware taint analysis for android apps.
\newblock In {\em PLDI}, 2014.

\bibitem{backstage}
V.~Avdiienko, K.~Kuznetsov, I.~Rommelfanger, A.~Rau, A.~Gorla, and A.~Zeller.
\newblock Detecting behavior anomalies in graphical user interfaces.
\newblock In {\em ICSE}, 2017.

\bibitem{contextualintegrity}
A.~Barth, A.~Datta, J.~C. Mitchell, and H.~Nissenbaum.
\newblock Privacy and contextual integrity: Framework and applications.
\newblock In {\em S\&P}, 2006.

\bibitem{bianchi2015app}
A.~Bianchi, J.~Corbetta, L.~Invernizzi, Y.~Fratantonio, C.~Kruegel, and
  G.~Vigna.
\newblock What the app is that? deception and countermeasures in the android
  user interface.
\newblock In {\em S\&P}, 2015.

\bibitem{borges2017data}
N.~P. Borges~Jr.
\newblock Data flow oriented ui testing: exploiting data flows and ui elements
  to test android applications.
\newblock In {\em ISSTA}, 2017.

\bibitem{peg}
K.~Z. Chen, N.~M. Johnson, S.~Dai, K.~MacNamara, T.~R. Magrino, E.~X. Wu,
  M.~Rinard, and D.~X. Song.
\newblock Contextual policy enforcement in android applications with permission
  event graphs.
\newblock In {\em NDSS}, 2013.

\bibitem{chen2015droidjust}
X.~Chen and S.~Zhu.
\newblock Droidjust: automated functionality-aware privacy leakage analysis for
  android applications.
\newblock In {\em WiSec}, 2015.

\bibitem{felt2011android}
A.~P. Felt, E.~Chin, S.~Hanna, D.~Song, and D.~Wagner.
\newblock Android permissions demystified.
\newblock In {\em CCS}, 2011.

\bibitem{fernandes2016appstract}
E.~Fernandes, O.~Riva, and S.~Nath.
\newblock Appstract: on-the-fly app content semantics with better privacy.
\newblock In {\em MobiCom}, 2016.

\bibitem{leaksemantic}
H.~Fu, Z.~Zheng, S.~Bose, M.~Bishop, and P.~Mohapatra.
\newblock Leaksemantic: Identifying abnormal sensitive network transmissions in
  mobile applications.
\newblock In {\em INFOCOM}, 2017.

\bibitem{flowintent}
H.~Fu, Z.~Zheng, A.~K. Das, P.~H. Pathak, P.~Hu, and P.~Mohapatra.
\newblock Flowintent: Detecting privacy leakage from user intention to network
  traffic mapping.
\newblock In {\em SECON}, 2016.

\bibitem{androido}
Google.
\newblock Android o behavior changes.
\newblock \url{https://developer.android.com/preview/behavior-changes.html},
  2017.

\bibitem{gorla2014checking}
A.~Gorla, I.~Tavecchia, F.~Gross, and A.~Zeller.
\newblock Checking app behavior against app descriptions.
\newblock In {\em ICSE}, 2014.

\bibitem{supor}
J.~Huang, Z.~Li, X.~Xiao, Z.~Wu, K.~Lu, X.~Zhang, and G.~Jiang.
\newblock Supor: precise and scalable sensitive user input detection for
  android apps.
\newblock In {\em USENIX Security}, 2015.

\bibitem{asdroid}
J.~Huang, X.~Zhang, L.~Tan, P.~Wang, and B.~Liang.
\newblock Asdroid: detecting stealthy behaviors in android applications by user
  interface and program behavior contradiction.
\newblock In {\em ICSE}, 2014.

\bibitem{peruim}
Y.~Li, Y.~Guo, and X.~Chen.
\newblock Peruim: understanding mobile application privacy with permission-ui
  mapping.
\newblock In {\em Ubicomp}, 2016.

\bibitem{miettinen2014conxsense}
M.~Miettinen, S.~Heuser, W.~Kronz, A.-R. Sadeghi, and N.~Asokan.
\newblock Conxsense: automated context classification for context-aware access
  control.
\newblock In {\em Asia CCS}, 2014.

\bibitem{smarper}
K.~Olejnik, I.~I. Dacosta~Petrocelli, J.~C. Soares~Machado, K.~Huguenin, M.~E.
  Khan, and J.-P. Hubaux.
\newblock Smarper: Context-aware and automatic runtime-permissions for mobile
  devices.
\newblock In {\em S\&P}, 2017.

\bibitem{pandita2013whyper}
R.~Pandita, X.~Xiao, W.~Yang, W.~Enck, and T.~Xie.
\newblock Whyper: Towards automating risk assessment of mobile applications.
\newblock In {\em USENIX Security}, 2013.

\bibitem{ringeraudacious}
T.~Ringer, D.~Grossman, and F.~Roesner.
\newblock Audacious: User-driven access control with unmodified operating
  systems.
\newblock In {\em CCS}, 2016.

\bibitem{ross2008incremental}
D.~A. Ross, J.~Lim, R.-S. Lin, and M.-H. Yang.
\newblock Incremental learning for robust visual tracking.
\newblock {\em International journal of computer vision}, 77(1):125--141, 2008.

\bibitem{xposed}
rovo89.
\newblock Xposed.
\newblock
  \url{http://repo.xposed.info/module/de.robv.android.xposed.installer}, 2017.

\bibitem{virusshare}
Virusshare.
\newblock Virusshare.
\newblock \url{https://virusshare.com/}, 2017.

\bibitem{weka}
waikato.
\newblock Weka.
\newblock \url{http://www.cs.waikato.ac.nz/ml/weka/}, 2017.

\bibitem{wagner2015}
P.~Wijesekera, A.~Baokar, A.~Hosseini, S.~Egelman, D.~Wagner, and K.~Beznosov.
\newblock Android permissions remystified: A field study on contextual
  integrity.
\newblock In {\em USENIX Security}, 2015.

\bibitem{wagner2017}
P.~Wijesekera, A.~Baokar, L.~Tsai, J.~Reardon, S.~Egelman, D.~Wagner, and
  K.~Beznosov.
\newblock The feasibility of dynamically granted permissions: Aligning mobile
  privacy with user preferences.
\newblock In {\em S\&P}, 2017.

\bibitem{appintent}
Z.~Yang, M.~Yang, Y.~Zhang, G.~Gu, P.~Ning, and X.~S. Wang.
\newblock Appintent: Analyzing sensitive data transmission in android for
  privacy leakage detection.
\newblock In {\em CCS}, 2013.

\bibitem{yin2016incremental}
X.~Yin, W.~Shen, and X.~Wang.
\newblock Incremental clustering for human activity detection based on phone
  sensor data.
\newblock In {\em CSCWD}, 2016.

\bibitem{zhang2016rethinking}
Y.~Zhang, M.~Yang, G.~Gu, and H.~Chen.
\newblock Rethinking permission enforcement mechanism on mobile systems.
\newblock {\em IEEE Transactions on Information Forensics and Security},
  11(10):2227--2240, 2016.

\end{thebibliography}

\end{document}